\documentclass{aastex}
\usepackage{emulateapj5}
\usepackage{psfig}
\begin{document}

%\documentclass[preprint]{aastex}

% \tightenlines
% \mathwithsecnums
 \received{21 May 2002}
 \accepted{13 Jun 2002}
% \journalid{}
% \articleid{}{}

%\lefthead{}
%\righthead{}
\lefthead{Kundu et al.}
\righthead{Low Mass X-ray Binaries in NGC 4472}

\title{The Low-Mass X-ray Binary-Globular Cluster connection in NGC 4472\altaffilmark{1}}

\author{Arunav Kundu\altaffilmark{2}, Thomas J. Maccarone\altaffilmark{3} \& Stephen E. Zepf\altaffilmark{2} }

\altaffiltext{1} {Based on observations made with the NASA/ESA Hubble Space Telescope, obtained at the Space Telescope Science Institute, which is operated by
    the Association of Universities for Research in Astronomy, Inc., under NASA contract
    NAS 5-26555, and on observations made with the Chandra X-ray Observatory.}
\altaffiltext{2}{ Physics \& Astronomy Department, Michigan State University, East Lansing, MI 48824. e-mail: akundu, zepf @pa.msu.edu}

\altaffiltext{3}{ SISSA/ISAS, via Beirut n. 2-4, 34014, Trieste, Italy. e-mail: maccarone@ap.sissa.it}

\begin{abstract}
	We have analyzed the low mass X-ray binary (LMXB) candidates in a {\it Chandra} observation of the giant elliptical galaxy NGC 4472. In a region observed by the Hubble Space Telescope (HST),  approximately 40\% of the bright (L$_X$$\gtrsim$10$^{37}$ ergs s$^{-1}$) LMXBs are associated with
optically identified globular clusters (GC). This is significantly higher than the fraction of bright LMXBs in  Galactic GCs and confirms that GCs are the dominant sites of LMXB formation in early type galaxies. The
 $\approx$4\% of NGC 4472 GCs hosting bright LMXBs, on the other hand, is remarkably similar
to the fraction of GCs with LMXBs in every other galaxy. 
Although statistical tests suggest that the luminosity of a cluster is an important driver of LMXB formation in GCs, this appears largely to be a
consequence of the greater number of stars in bright clusters. The metallicity
of GCs is a strong determinant of LMXB specific frequency, with  metal-rich clusters about 3 times more likely to host  LMXBs than metal-poor ones. There are weaker dependences on the size of a GC and its distance from the center of the galaxy. The X-ray luminosity does not depend significantly on the properties of the host GC.
Furthermore, the spatial distribution and X-ray luminosity function of LMXBs within and outside GCs are indistinguishable. The X-ray luminosity function of {\it both} GC-LMXBs and non-GC-LMXBs reveal a break at $\approx$3$\times$10$^{38}$ ergs s$^{-1}$ 
strongly suggesting that the brightest LMXBs are black hole accretors. 
 
\end{abstract}


\section{Introduction}
	
	Early X-ray studies with the {\it Einstein} X-ray observatory
revealed that elliptical  and S0 galaxies are significant sources of X-ray emission. While the bulk of the soft X-ray luminosity of X-ray bright galaxies can be attributed to kT $\sim$ 1 keV thermal emission from hot (10$^7$ K) gas (e.g. Forman, Jones \& Tucker 1985), the hard X-ray flux is roughly proportional to the optical luminosity, suggesting a low-mass X-ray binary origin (Trinchieri \& Fabbiano 1985). With the advent of the {\it Chandra} X-ray Observatory, it is
 now possible to resolve the point source X-ray populations likely to be associated with LMXBs in external galaxies such as NGC 4697 (Sarazin, Irwin \& Bregman 2000) and NGC 1399 (Angelini, Loewenstein \& Mushotzky 2001; hereafter A01), and definitively show that LMXBs make a significant contribution to the X-ray flux in typical ellipticals. Since the stellar populations in such galaxies are at least a few Gyrs old (e.g. Trager et al. 2000), contamination by
high mass X-ray binaries is not an issue.

	Globular clusters are especially fertile environments for LMXB formation.  Even though GCs account for $\lesssim$0.1\% of the stellar mass in the Galaxy, they harbor about 10\% of the L$_X$$\gtrsim$10$^{36}$ erg s$^{-1}$ LMXBs (e.g. Verbunt 2002), indicating  a probability of hosting a LMXB that is at least two orders of magnitude larger than for field stars. Recent {\it Chandra} observations indicate that at least 20\% of the LMXBs in NGC 4697 (Sarazin, Irwin, \& Bregman 2001) and as much as 70\% in NGC 1399 (A01) may reside 
in GCs, suggesting that GCs are even more dominant sites of LMXB
formation in galaxies without recent star formation. 

 Dynamical processes, such as tidal capture of neutron stars  in close encounters with other cluster 
stars (Clark 1975; Fabian et al. 1975), and interactions between single stars and 
binaries (Hills 1976) in the high density environment of GC cores are some of the leading explanations for the preponderance of
 LMXBs in GCs. However, the very small number (13) of bright LMXBs in Galactic
globulars (e.g. Liu, van Paradijs \& van den Heuvel 2001; hereafter L01]) and other local group GCs has been the biggest impediment
in testing these models in any detail. 
Thus, in order to isolate
and understand the primary environmental factors driving LMXB formation in GCs
 it is imperative to study them in distant galaxies, especially globular cluster rich
early types. In this Letter we analyze the LMXB population of 
NGC 4472, the giant elliptical in Virgo, and its connection to the GC system.
 
\section{Observations \& Data Reduction}

We  have analyzed the $\approx$40 ks archival {\it Chandra} ACIS-S3 image of NGC 4472 obtained by Mushotzky et al. on June 12, 2000. After standard data pipeline processing, as described on the PSU
 webpage \footnote {http://www.astro.psu.edu/xray/acis/recipes/clean.html}, we
identified the X-ray point source populations in the 0.5-2.0 keV, 2.0-8.0 keV, and 0.5-8 keV images. Using WAVDETECT from the CIAO 2.2 package, with a threshold of 10$^{-6}$ probability of false
detections ($\lesssim$1 false source per field) we detected 148 total sources.  Of these we eliminate four that are either within 8$''$ of the nucleus and may be associated either with the central AGN or density enhancements in the hot interstellar gas, or have  shapes inconsistent with point sources. About 10-15 of the 144 candidates are likely to be associated with contaminants such as background AGNs (Brandt et al. 2000, Mushotzky et al. 2000); the bulk of the rest are expected to be LMXBs. The hardness ratios (Counts[2-8 keV])/Counts[0.5-2 keV]) of all the sources are consistent with a value of 0.38 at the 2$\sigma$ level and cannot be used to distinguish
GCs from contaminants, in agreement with the findings of A01. We extract spectra with the PSEXTRACT script
using all channels from 0.5 - 8 keV. Freezing the neutral hydrogen column density of $1.3\times 10^{20}$ cm$^{-2}$ we fit a power law spectra in XSPEC 11.0 (Arnaud 1996). A distance of 16 Mpc, a
 typical value for the Virgo cluster (e.g. Macri et al. 1999), is adopted to convert the fluxes to luminosity,
and for all other calculations in this Letter.  The faintest LMXB candidate detected has a luminosity of 1.1$\times$10$^{37}$ ergs s$^{-1}$ in the 0.5 - 8 keV band in which we measure L$_X$.   Detailed reduction procedures and
 source lists will be presented in a companion paper (Maccarone, Kundu, \& Zepf 2002).

	In addition to the central HST-WFPC2 image of NGC 4472 previously 
studied by us (Kundu \& Whitmore 2001; hereafter KW01) we  analyzed 3 other sets of archival
 WFPC2 V \& I images of the galaxy and its halo (Fig. 1). Using the techniques described in KW01 we identified 825 GC candidates in the color range 0.8$<$ V-I $<$1.4 within the ACIS-S3 image, excluding the region within 10$''$ of the nucleus, and measured their luminosity, color and half-light radii (r$_h$). Note
 that we used updated zero points from the HST Data Handbook V 3.1 and re-calculated aperture corrections for all the pointings for this analysis. We detect $>$80\% of the total GC population within the ACIS-WFPC2 overlap region, with an estimated contamination fraction of a few percent (KW01). Given the high density of GCs, sub-arcsecond relative astrometry is essential to eliminate false matches. We achieve 0.3$''$ r.m.s. relative astrometric accuracy by bootstrapping the {\it HST} and {\it Chandra} positions of obvious matches to
the ground-based Mosaic image of Rhode \& Zepf (2001).

\section{Results \& Discussion}

	Even though the WFPC2 images cover only $\approx$20\% of the ACIS-S3 frame, 72 of the 144 LMXB candidates lie within this  region. We consider LMXBs to be associated with GCs if they are separated by less than 0.65$''$, where there is a natural break in the LMXB-GC angular separation. Thirty LMXBs satisfy this criterion. We note that increasing the matching distance to 1$''$ only adds 1 more source, while removing the color criterion adds 2 candidates with colors inconsistent with GCs. We choose not to relax the matching criterion in order to minimise the possibility of spurious matches.	 Thus, 40\% of 
the L$_X$$>$10$^{37}$ ergs s$^{-1}$ LMXBs detected in NGC 4472 lie in GCs. This is significantly higher than the 5-15\% figure in the Milky Way - based on the L01 catalog and Harris (1996) GC distances, with the variation due to how transient sources are counted - but smaller than the $\approx$70\% measured in NGC 1399 (A01).  The fraction of LMXBs in GCs and other associated numbers reported here represent ``snapshot" values at the observational epoch. While it is possible that there 
are underlying populations of long-lived transients in the field or GCs, the
 instantaneous values are central to addressing scientifically interesting
 issues such as the importance of LMXBs to the X-ray emission from galaxies, their effect on X-ray background measurements and the significance of GCs to these values. Hence it appears that at present GCs are the dominant
sites for LMXB formation in early type galaxies, in consonance with the suggestion of White, Sarazin \& Kulkarni (2002).  Thus,  it is vitally important to study the GC-LMXB connection in these galaxies.

	Although the fraction of LMXBs in GCs is different across galaxy types, the fraction of GCs hosting LMXBs is remarkably similar in all galaxies studied
to date. Approximately 4\% of NGC 4472 GCs host L$_X$$\gtrsim$10$^{37}$ erg s$^{-1}$ LMXBs, compared to about 4\% in NGC 1399 (A01),  $\approx$2-3\% in M31 (using DiStefano et al. 2002 and Barmby \& Huchra 2001), and 1-4\% in the Galaxy (using Liu et al. 2001 and Harris 1996).
Thus the formation efficiency of LMXBs in GCs must be strongly driven by the
properties of GCs, rather than that of the host galaxy. 

\subsection{Which Clusters Preferentially Form LMXBs?}
	The color distribution of  the NGC 4472 GCs is clearly bimodal (Fig. 2),
with a population of blue, metal-poor GCs, and one of red, metal-rich GCs, as
found in most other ellipticals (e.g. Zepf \& Ashman 1993, KW01). KMM mixture modeling tests (Ashman, Bird \& Zepf
 1994) reveal a blue peak at V-I=0.98 and a red one at V-I=1.23, with a dividing color of V-I=1.10. One of the more striking results (Fig. 2) is that there are 3.3 times as many LMXBs in red GCs as there are in the
 blue. Even after accounting for the slightly larger number of red GCs, the
overabundance factor is still $\approx$2.7. Previous observations of Galactic GCs have also showed such a trend (e.g. Bellazzini et al. 1995), although 
small number statistics make it difficult to disentangle it from other
GC properties such as distance from the center of the Galaxy.  

The top panel of Fig. 2 also suggests a  tendency for more efficient LMXB formation in GCs in the inner region. The lower 
panel of Fig. 2 clearly shows that LMXBs are formed preferentially in the brightest GCs. This effect is actually even stronger than indicated when one considers that the faint end of the globular cluster luminosity function (GCLF) has not been corrected for incompleteness. Figure 2 also 
suggests that smaller GCs are favored LMXB hosts. 
Thus, each of these variables could provide the physical environment that promotes
LMXB formation in GCs. In order to better understand the relative importance of each of these factors we turn to discriminant analysis (DA).

	DA is used to
 weight and combine the discriminating variables in such a way that
the differences between pre-defined groups are maximized (e.g. Antonello \& Raffaelli 1983). Thus each data point is assigned a discriminant score of the form $F = w_1x_1 + w_2x_2 +  ... w_ix_i$
where F is the discriminant score, $w_i$ is the weighting coefficient for variable i, and $x_i$ is the i$^{th}$ discriminating variable, such that the
distribution of discriminant scores of the pre-defined groups is maximally
separated along the axis of this new composite variable. The absolute values of
 the standardized  coefficients,  $w_i$,  reveal the relative
importance of the associated discriminating variables. In certain cases, where the discriminating variables may be correlated,
the absolute value of the structure coefficients - which are the correlations of
each variable with the discriminant function - may give better estimates of the 
significance of each of the variables. 

Using the SPSSv10 package we performed DA on the LMXB and non-LMXB GC populations with V, V-I, r$_h$ and distance from the center of the galaxy as the variables. The standardized coefficients and structure coefficients (within brackets) for selected tests are presented in Table 1.  Since DA is a linear method, we tested the importance of GC size by using r$_h$, r$_h^3$ and log(r$_h$) in turn. We report only the r$_h$ analysis, which consistently assigns the
most power to the size parameter. Two random variables, a Gaussian, and a uniform distribution are also used to gauge the significance of the results. Fig. 3 shows the success of DA in separating the two populations. DA of the entire sample clearly shows that the luminosity of a GC is  the 
most important factor that drives LMXB formation in GCs.  Color and distance appear to be equally important secondary factors that drive LMXB
formation, with size providing a much weaker discriminant. We note however that the incompleteness of the faint end of the GCLF has a radial dependance that is both a function of the background light and differences in the exposure times of the WFPC2 images. Since the GCLF is constant at
all radii (Kundu, Zepf \& Ashman 2002) and the luminosity is known to be uncorrelated with color and r$_h$, restricting the sample to V$<$23.5
mag GCs, where the completeness is $\approx$100\%, provides a fairer statistical test for distance effects. 
This sample reveals that V, and V-I are the two most important discriminating variables, followed by smaller contributions from r$_h$ and distance. An independent statistical test using logistic regression analysis  reveals similar weights for each parameter, strengthening our confidence in the results.
 
	The strong dependence of LMXB specific frequency on GC luminosity could    simply be
 a reflection of the larger number of stars in more luminous GCs, or it could possibly signal more efficient dynamical LMXB formation in massive GCs. The inset in Fig. 3 shows that the LMXB density per unit GC luminosity is roughly constant with V magnitude. Since
GC luminosity is proportional to the total number of stars for the roughly constant M/L (and similar IMFs) in GCs , this suggests that
the dependence of LMXB frequency on luminosity can largely be attributed to the
greater number of stars in bright GCs.  It is also clear from our analysis that metallicity is a significant,
strong, independent parameter. There are few theoretical explanations that can account for this. One possibility is the suggestion of Bellazzini et al (1995), that the larger radii of metal-rich stars promotes LMXB formation in GCs and facilitates mass transfer by Roche lobe overflow. While Grindlay (1987) suggested that a similar metallicity trend seen in Galactic GCs can be explained by a flatter initial mass function in metal-rich GCs, recent studies suggest that the observed differences in the present day mass function depend primarily on GC evolution, and not metallicity (e.g. Piotto \& Zoccali 1999). While the rate of LMXB formation depends on the GC environment, our observations reveal no obvious trend of  L$_X$ with GC luminosity, color, size or distance. Further theoretical studies of the effect
of GC environment on LMXB formation and evolution are required to understand the underlying physics.

\subsection{LMXBs in Clusters vs. LMXBs in the Field} 
	Are LMXBs in GCs different from those outside GCs? Were the non-GC-LMXBs in ellipticals formed in GCs and consequently ejected? In
order to address some of these questions we compare the GC and non-GC populations of LMXBs. The mean luminosity of GC-LMXBs, $<L_X>=2.4\pm0.6 \times 10^{38}$ erg s$^{-1}$ is slightly higher than, but statistically indistinguishable from, the non-GC-LMXBs at $<L_X>= 1.8\pm0.3 \times 10^{38}$ erg s$^{-1}$.  Similarly, Kolmogorov-Smirnov tests revealed that the possibility that the GC and non-GC-LMXBs are drawn from the same L$_X$ or radial distance populations cannot be eliminated at greater than the 30\% level. However, the exact distribution of each of these quantities in the
two populations may hold vital clues about their origin. A single power-law fit  of the cumulative luminosity function (Fig. 4) of each of the LMXB categories can be eliminated at greater than 
 95\% confidence level, while a broken
power law of the form $N(>L) = N_{cut}(\frac{L}{L_{cut}})^{\alpha_{above/below}}$
 fits each of the populations well. The non-GC-LMXBs and ``All" 
source lists were corrected for background/foreground contamination, estimated from Mushotzky et al (2000), before fitting.  As is evident from Fig. 4, all populations show a break at L$_{cut}$$\approx$3$\times$10$^{38}$ ergs s$^{-1}$.
 A similar break is observed in other galaxies; Sarazin et al (2000) argue that the ``knee" is at the typical Eddington luminosity for spherical accretion onto a 1.4 M$_\odot$ neutron star, and may indicate the break between
neutron star and black hole (BH) accretors. The luminous GC-LMXB population cannot be explained by multiple bright LMXBs in favorable GC environments as we 
see no trend of higher X-ray luminosity in preferred GC hosts. Also, the  ``knee" in the non-GC-LMXB population of NGC 4472 cannot be caused by multiple LMXBs. Thus BH-LMXBs appear to be the most plausible explanation for the break based on present observations.

 If non-GC-LMXBs were formed in clusters they can escape from GCs by one of
two mechanisms, cluster destruction  - which is more efficient in the inner regions of galaxies (e.g. Vesperini 2000), and should reveal a surfeit of non-GC-LMXBs in the inner regions - or dynamical kicks that eject LMXBs from GCs. In the latter case, one might expect a spatial density distribution at least as extended
as that of the GCs. If non-GC-LMXBs are formed in the field population, one would of course expect the density distribution to follow the light profile (which is steeper than the GC profile). The bottom panel of Fig. 4 suggests that the radial density distribution 
of GC-LMXBs and non-GC-LMXBs is roughly similar, and consistent with the GC population. At face value this suggests that dynamically ejected LMXBs from GCs
may account for a significant fraction of non-GC-LMXBs.  Studies of the spatial properties of LMXBs over a larger radial range, especially the innermost regions of X-ray faint galaxies, are required to probe whether most non-GC-LMXBs in early type galaxies have a GC ancestry. Second epoch {\it Chandra} observations
will further allow discrimination between possible long-lived transient LMXBs
associated with the field, and the more persistent sources that may be expected to form in GCs (Piro \& Bildsten 2002).

	SEZ and AK gratefully acknowledge support from NASA via the LTSA grant
NAG5-11319. We thank Enrico Vesperini for many useful discussions.

\begin{figure}
\plotone{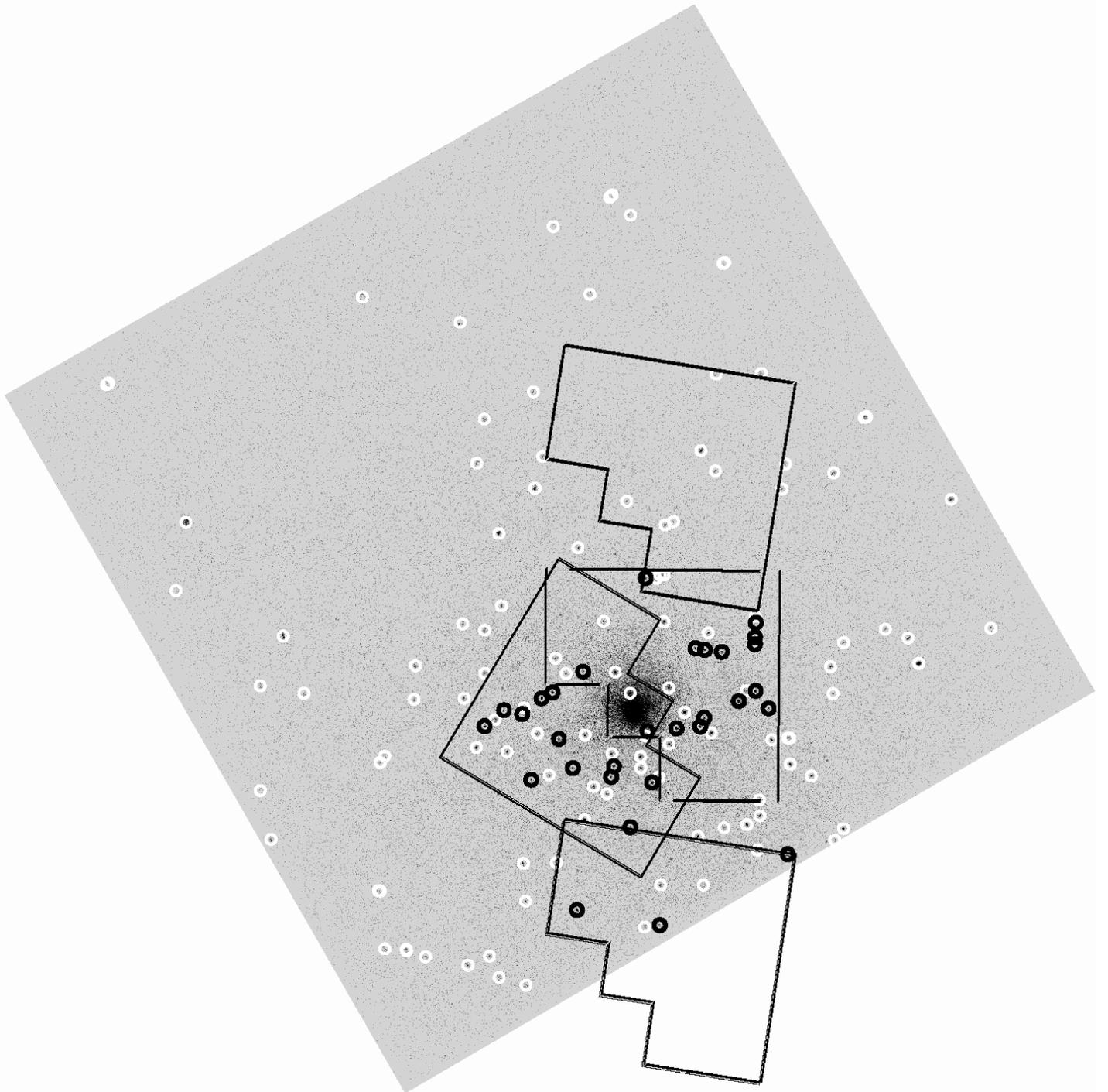}
%\plotfiddle{fig1.epsi}{6in}{0}{100.}{100.}{-300}{-100}
%\centerline{\psfig{figure=fig1.epsi,width=15cm,angle=0}}
\caption{Chandra ACIS-S3 image of NGC 4472 in the 0.5-8.0 keV band with the HST-WFPC2 fields-of-view superposed. Light circles mark LMXB candidates. Dark circles indicate the positions of GC matches.}
\end{figure}

\begin{figure}
\plotone{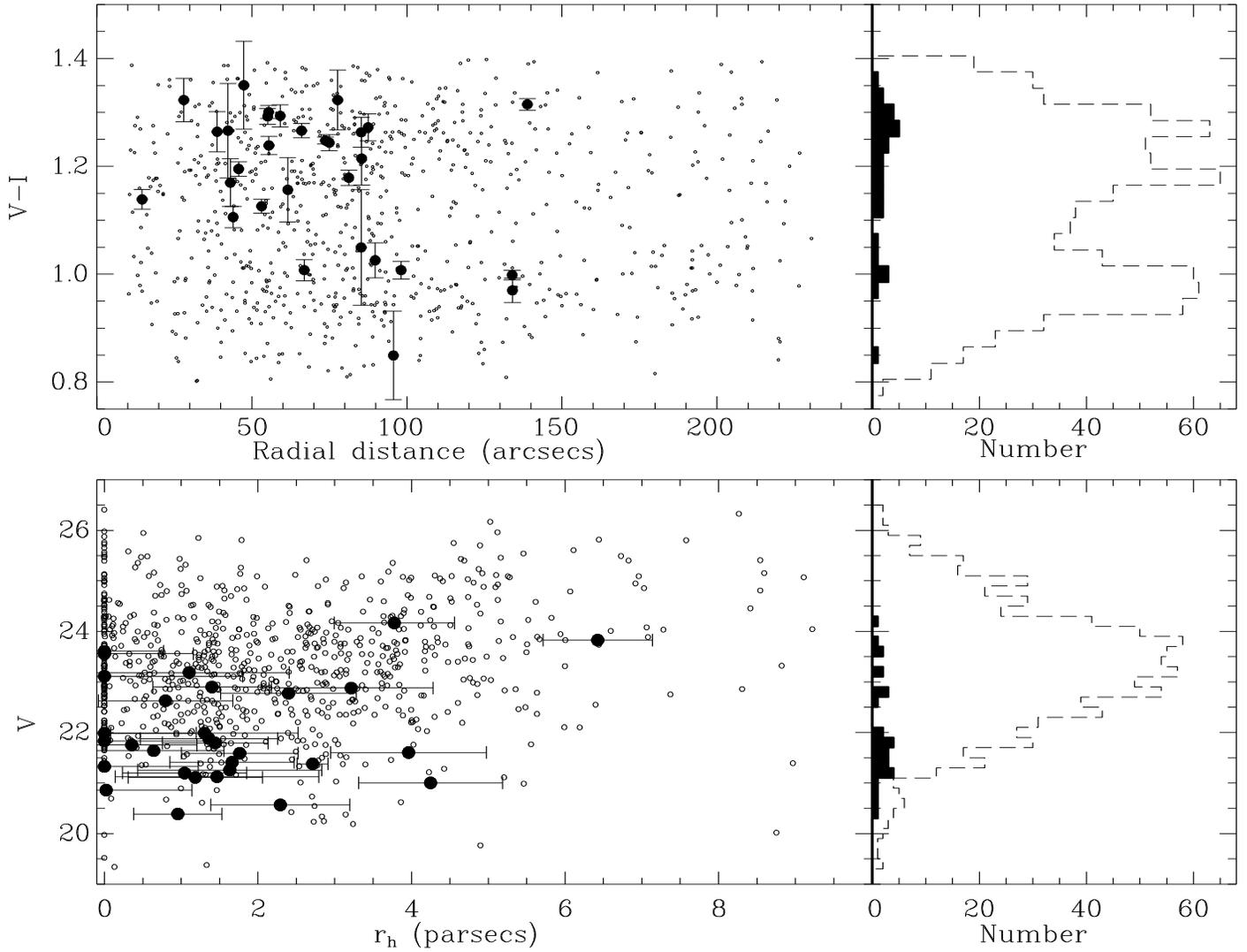}
%\centerline{\psfig{figure=fig2.epsi,width=15cm,angle=0}}
\caption{Top: The V-I colors of GCs vs. distance from the center of NGC 4472 and GC color distribution. LMXB-GC matches are indicated by filled circles/bins.  Bottom: V magnitude of globular clusters vs. half light radius and the GCLF. }
\end{figure}

\begin{figure}
\plotone{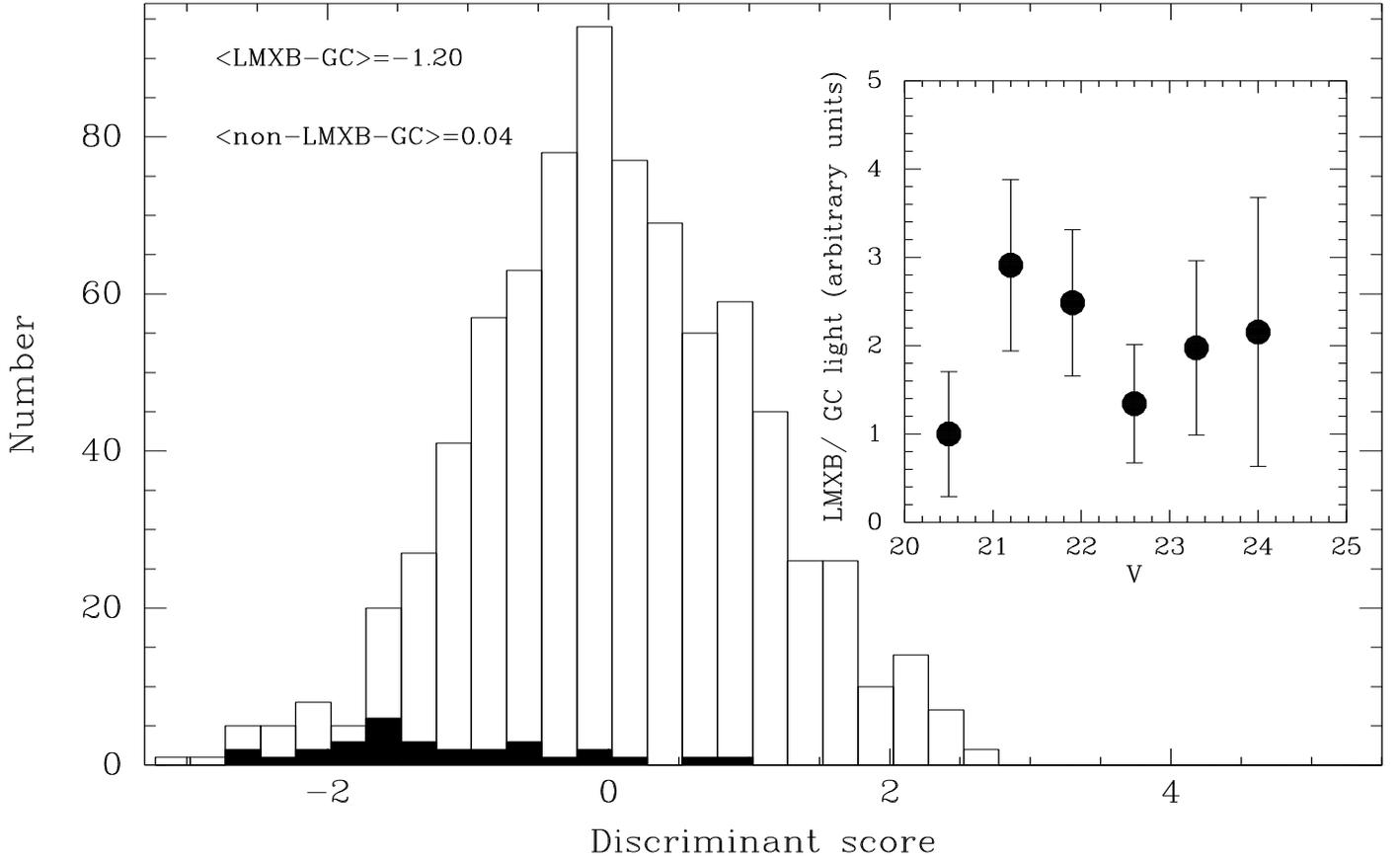}
%\centerline{\psfig{figure=fig3.epsi,width=15cm,angle=0}}
\caption{Representative histograms of the discriminant scores of LMXB-GCs and non-LMXB GCs from discriminant analysis using V, V-I, distance, and r$_h$ as discriminating variables. The two populations are well separated. Inset shows the LMXB efficiency per unit GC light in arbitrary units as a function of V magnitude.}
\end{figure}

\begin{figure}
\epsscale{0.6}
\plotone{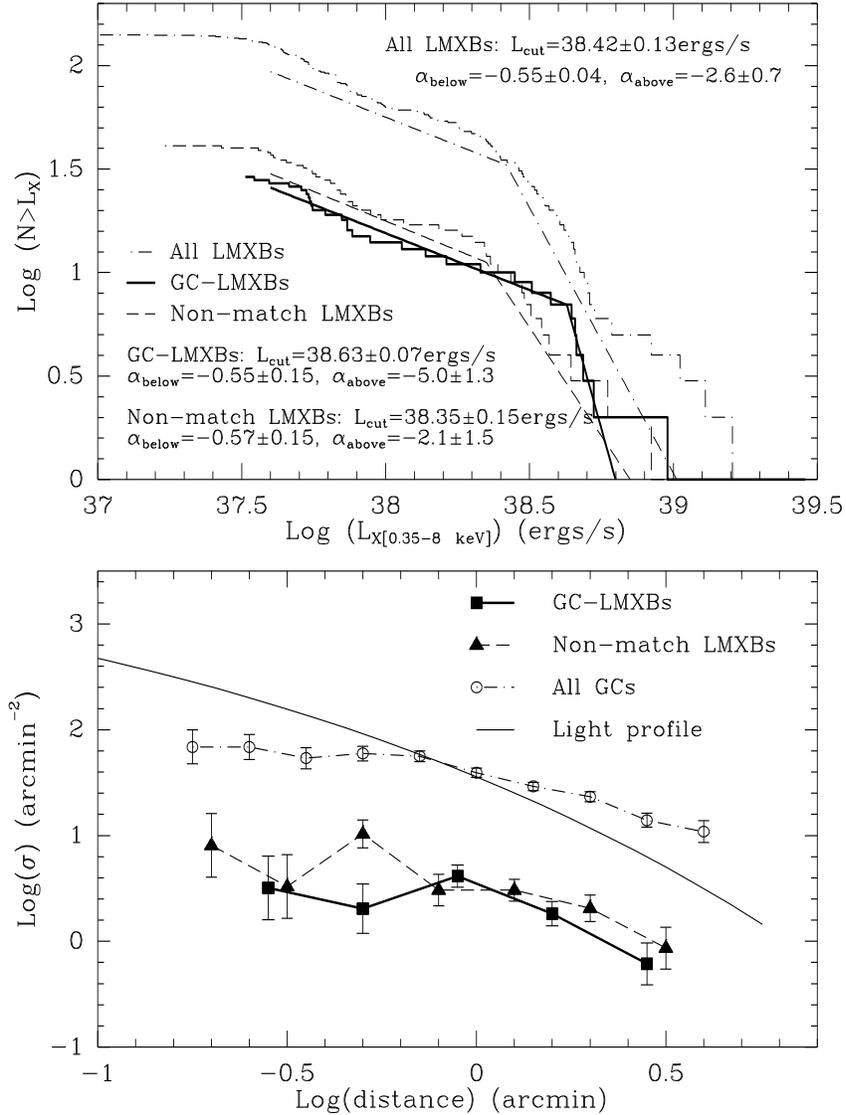}
%\centerline{\psfig{figure=fig4.epsi,width=15cm,angle=0}}
\caption{Top: Histogram of the cumulative X-ray luminosity functions. Note that  the histograms  have {\it not} been corrected for background/foreground contaminants. Lines indicate broken power-law fit to the GC-LMXB profile and contamination corrected fits to the non-GC-LMXB and all LMXB profiles. A second possible break at $\approx$10$^{37.5}$ ergs s$^{-1}$ may be due to incompleteness effects, or reflect the counterpart of the break seen in the Galaxy and M31.  Bottom: Radial surface density profile of GC and non-GC-LMXBs with Poisson error bars. The GC profile is derived from the density of V$<$23.5 mag GCs  (selected so that the completeness is $\approx$100$\%$ everywhere) and shifted by
an arbitrary amount vertically. The light profile from Rhode \& Zepf (2001), shifted by an arbitrary amount, is shown for comparison.  }
\end{figure}

\begin{deluxetable}{lllllllllll}
\tabletypesize{\small}
\tablenum{1}
\tablecaption{Discriminant Analysis Weights }
\startdata
    \\
\tableline
\tableline
V   		&  V-I 		& Distance	& r$_h$ 		& Random1\tablenotemark{1} 	& Random2\tablenotemark{1} 		 & p-value\tablenotemark{2} 	& Correct\tablenotemark{2}\\
								
%out_all												
0.89(0.89)	& -0.41(-0.35)& 0.14(0.34)	& 0.04(0.22)		& -0.04(-0.07)	& 0.05(0.10)		& 0.000	& 74.3\% \\
%var_all												
0.89(0.90)	& -0.41(-0.35)& 0.14(0.34)	& 0.04(0.22)		& 		&	 		& 0.000	& 74.4\% \\

\multicolumn{8}{c}{\underline{BRIGHT GLOBULAR CLUSTERS (V$<$23.5)}}\\		
%var_all_bright												
0.86(0.82)	& -0.43(-0.40) & 0.19(0.23)	& 0.28(0.30)		& 		& 			& 0.000	& 75.5\% \\

\tableline		 

\enddata
\tablenotetext{1} {Random1 and Random2 are dummy Normal and uniform random variables respectively} 
\tablenotetext{2} {The ``p-value" denotes the significance of the null hypothesis that there is no discriminating power in the variables based on Wilk's lambda statistic, while ``Correct" reports the percentage of cases classified in the right group by DA.} 
\end{deluxetable}

\end{document}